\documentclass[aps,prb,twocolumn,a4paper,superscriptaddress,floatfix,showpacs,preprintnumbers,amsmath,amssymb]{revtex4-2}
\usepackage{url}
\usepackage[english]{babel}
\usepackage[utf8]{inputenc}
\usepackage{amsmath}
\usepackage{graphicx,epstopdf}
\usepackage{amssymb,bm,epsfig,color,textcase}
\usepackage{hyperref}
\providecommand{\U}[1]{\protect\rule{.1in}{.1in}}

\begin{document}


\title{Proximity-induced sequence of field transitions in Kitaev candidate BaCo$_2$(AsO$_4$)$_2$}

\author{Pavel A. Maksimov}%
\affiliation{Bogolyubov Laboratory of Theoretical Physics, Joint Institute for Nuclear Research, Dubna, Moscow region 141980, Russia}
\affiliation{M.N. Miheev Institute of Metal Physics of Ural Branch of Russian Academy of Sciences, S. Kovalevskaya St. 18, 620990 Ekaterinburg, Russia}
\email{maksimov@theor.jinr.ru}

\date{\today}

\begin{abstract}
We study field-induced phase transitions of the minimal exchange model proposed earlier for BaCo$_2$(AsO$_4$)$_2$, a candidate for Kitaev honeycomb model, using numerical minimization of classical spin clusters. We show that experimentally observed sequence of step-like transitions in magnetic field is realized in the phase diagram of the minimal model. Surprisingly, intermediate up-up-down plateau phase is stabilized only in the proximity of a double-zigzag$-$zigzag phase boundary. We systematically map out the region of stability of experimentally observed cascade of transitions and argue that BaCo$_2$(AsO$_4$)$_2$ exchange parameters are close to a region of strong phase competition, which can explain suppressed saturation field.
\end{abstract}

\pacs {75.10.Jm, 75.25.+z, 75.30.Kz}

\maketitle

{\it Introduction.-}  Kitaev $S=1/2$ honeycomb exchange model \cite{KITAEV2006} is exactly solvable and has a quantum spin liquid (QSL) ground state, a correlated state without long-range magnetic order \cite{balents_review,savary_balents}.  The model involves exchanges on three types of bonds of the honeycomb lattice with compass-like Ising interactions of three different spin axes \cite{nussinov_compass_2015}, and was shown to host exotic Majorana and vison excitations. The search for materials that can host QSL is invigorated by the promise of application of its excitations in topologically protected quantum computation \cite{Kitaev_toric_1998,kitaev2003fault,KITAEV2006,Nayak_review}.

Honeycomb $d^5$ Mott insulators with strong spin-orbit coupling and edge-sharing ligand octahedra were proposed by G. Jackeli and G. Khaliullin \cite{Jackeli} to host Kitaev honeycomb model. Subsequent experiments on $\alpha$-RuCl$_3$ \cite{plumb2014,banerjee2016} and Na$_2$IrO$_3$ \cite{Gegenwart_2012,Coldea12}, indeed, indicated the presence of Kitaev-type anisotropic exchanges, which was also supported by \textit{ab initio} calculations \cite{Katukuri,Perkins,valenti16,yadav2016,hou_2017,Berlijn19}. However, the existence of additional interactions allowed by symmetry of the lattice \cite{chaloupka_zigzag_2013,rau_jkg}, as well as sizeable third-neighbor exchanges, yields antiferromagnetic zigzag magnetic order in these materials. Nonetheless, the search for Kitaev interactions encouraged ongoing synthesis and analysis of new materials based on honeycomb \cite{Cava_RuI3,Ohgushi_RuBr3,Kim2021,Kaib2022,Cu2IrO3,Kitagawa2018,Winter_review,Takagi_review,Motome_2020,Tsirlin_review,Trebst_review} and triangular lattices \cite{Trebst_JK_triang,SciRep,Ioannis_Z2}.

\begin{figure}
\centering
\includegraphics[width=0.99\linewidth]{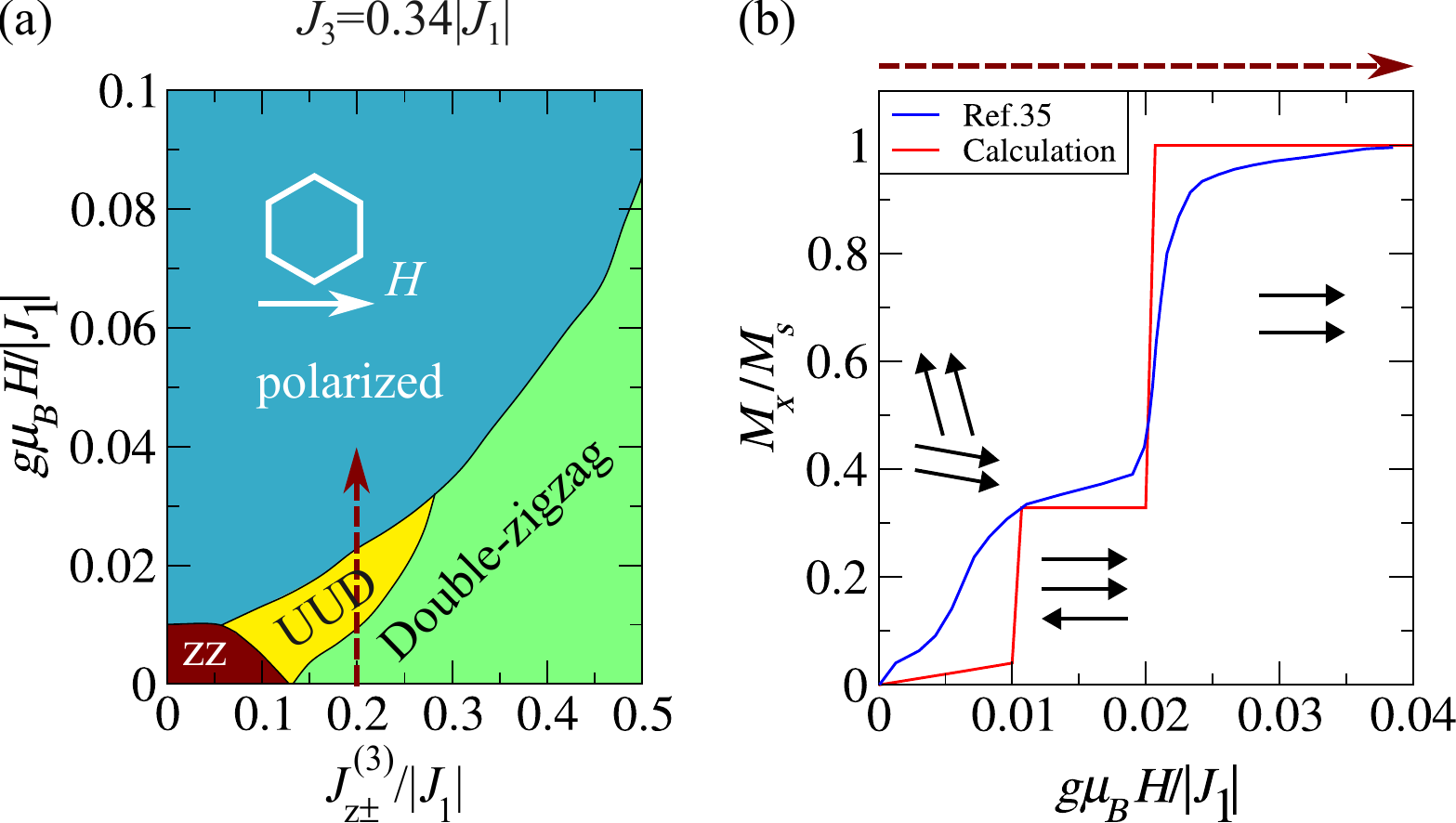}
\caption{A sketch of summarized results presented below in this paper. (a) Field phase diagram of the minimal model \eqref{eq_ham} for $J_3=0.34 |J_1|$ as a scan along $J_{z\pm}^{(3)}$, obtained with numerical minimization of classical spins.  Honeycomb lattice of BaCo$_2$(AsO$_4$)$_2$ with the direction of magnetic field used in Hamiltonian \eqref{eq_ham} are shown in the inset. (b) Magnetization along $x$-axis, calculated for $J_{z\pm}^{(3)}=0.2|J_1|$ with double-zigzag, up-up-down and polarized phases shown. Measurements from Ref.~\cite{Zhong2020} are also reproduced.}
\label{fig_0}
\end{figure}

BaCo$_2$(AsO$_4$)$_2$ (BCAO) was recently proposed as a platform for Kitaev honeycomb model \cite{Zhong2020} in a new class of compounds with $d^7$ Co$^{2+}$ ions, which were introduced as an alternative path towards Kitaev interactions \cite{Motome_Co_2018,Giniyat_Co_2018,Winter_Co_2022,Kee_Co_2023}. In Co-based materials spin-orbit coupling yields lowest effective doublet in the $L=1$, $S=3/2$ state, while additional hopping paths due to active $e_g$ orbitals were suggested to increase Kitaev exchange \cite{Khaliullin_3d_2020}. This mechanism was supported by experimental evidence in various honeycomb materials \cite{Liu_Co_review,Coldea_Co_2020,Cava_Co_2007,LCSO_2020,Songvilay_Co_2020,Park_Co_2020}, as well as triangular lattice Kitaev compounds \cite{Park_CoI2}.  

Early neutron-scattering experiments on BaCo$_2$(AsO$_4$)$_2$ indicated a spiral ground state with $\mathbf{Q}=(0.25,0)$ ordering vector \cite{Regnault_BaCoAsO_1977}, which is stabilized in the classical Heisenberg model by the competition of first- and third-neighbor interactions, $J_1$ and $J_3$. However, a more refined polarized neutron-diffraction data \cite{Regnault2018} pointed to a previously undetected double-zigzag ground state, which has the same periodicity but a collinear $\uparrow \uparrow \downarrow \downarrow$ structure. It was shown later that this enigmatic state can be either stabilized by quantum fluctuations in the narrow region of the isotropic $J_1$-$J_3$ model \cite{shengtao_j1j3}, or by extension of the Heisenberg Hamiltonian by third-neighbor bond-dependent Kitaev-type coupling \cite{BCAO_minimal}. 

Generally, candidate materials for Kitaev interactions can require a large exchange parameter phase space to describe their magnetic properties, and its experimental extraction of the couplings may become highly challenging, see as an example the story of $\alpha$-RuCl$_3$ \cite{us_PRR}. Field-induced behavior, especially if it involves multiple phase transitions, can provide extremely useful information for the investigation of magnetic Hamiltonian \cite{YZGO_2021}. In this paper, we address peculiar experimentally observed feature of BaCo$_2$(AsO$_4$)$_2$ - a field-induced sequence of phase transitions, which includes an up-up-down (UUD) $\uparrow \uparrow \downarrow$ $1/3$-plateau state \cite{Regnault_BaCoAsO_1977,Zhong2020}. This sequence of transitions can be also seen in the magnetic spectrum of BaCo$_2$(AsO$_4$)$_2$ in THz measurements \cite{Armitage_THz_2021,Wang_THz_2021}. Moreover, the saturation field appears to be strongly suppressed relative to order of exchange interactions: $g\mu_B H_s/|J_1| \simeq 0.03$. Using numerical minimization of classical spin clusters up to 144 sites, we study magnetic field phase transitions of a model proposed in Ref.\cite{BCAO_minimal} for BaCo$_2$(AsO$_4$)$_2$, and show that $1/3$-plateau appears in the cascade of field-induced phases only in the vicinity of zero-field $\mathbf{Q}=(1/3,0)$ state, which is intermediate between double-zigzag and zigzag states. Moreover, we map out a region, where parameters of the model can fit experimentally observed ranges of field-induced phases.

\begin{figure}
\centering
\includegraphics[width=0.99\linewidth]{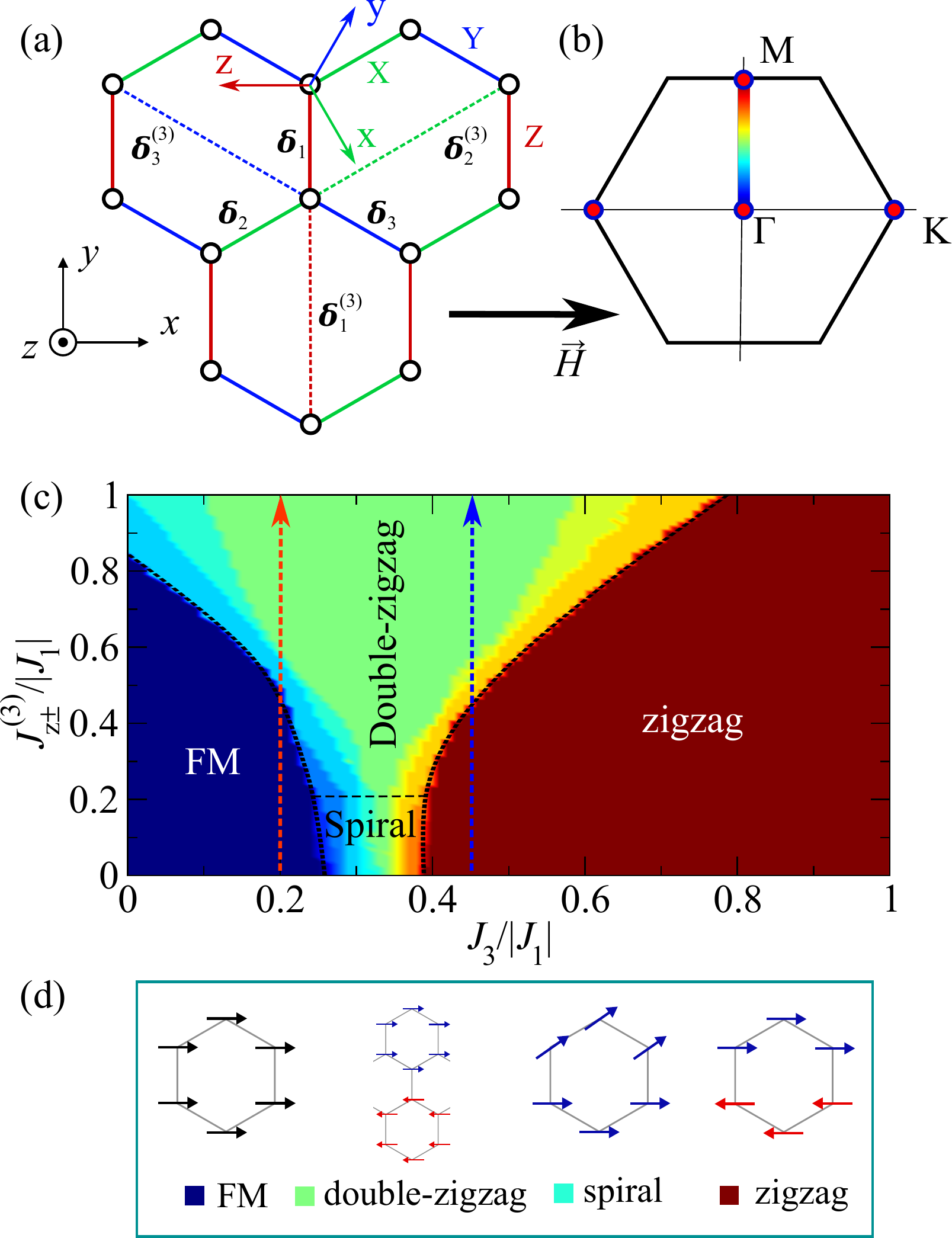}
\caption{(a) Honeycomb lattice with three types of nearest and third-neighbor bonds indicated with color. Crystallographic and cubic axes, $\{x,y,z\}$ and \{x,y,z\}, correspond to two representations of the extended Kitaev-Heisenberg model, see text. Direction of magnetic field is also shown. (b) Brillouin zone (BZ) of the honeycomb lattice with high-symmetry points marked. Colorbar illustrates the value of the ordering vector $\mathbf{Q}$ in the phase digram in (c). (c) Phase diagram of the model \eqref{eq_ham} in zero field, reproduced from Ref.\cite{BCAO_minimal}. Representative scans used in the Fig.~\ref{fig_scans} are shown with dashed arrows. (d) Sketches of primary phases in the phase diagram in (c).}
\label{fig_lattice_pd}
\end{figure}
{\it The model and field phase diagram.-} There are multiple attempts at establishing exchanges in BaCo$_2$(AsO$_4$)$_2$ both from neutron-scattering data \cite{Broholm_BCAO} and \textit{ab initio} calculations \cite{Paramekanti_Co_2021,BCAO_minimal}. We would like to argue that the model proposed in Ref.\cite{BCAO_minimal} sufficiently describes both a unique zero-field ground state and the peculiar sequence of field-induced phase transitions.

The exchange Hamiltonian used in this paper, which was coined as a ``minimal'' model for BaCo$_2$(AsO$_4$)$_2$ \cite{BCAO_minimal}, is given by
\begin{align}
\label{eq_ham}
\hat{\cal H}_\text{min}=-&g\mu_B H\sum_i S^x_i \\
+&\sum_{\langle ij\rangle_1}
J_1 \Big(S^{x}_i S^{x}_j+S^{y}_i S^{y}_j+\Delta_1 S^{z}_i S^{z}_j \Big) \nonumber\\
+& \sum_{\langle ij\rangle_3}J_3\Big(S^{x}_i S^{x}_j+S^{y}_i S^{y}_j +\Delta_3 S^{z}_i S^{z}_j\Big) \nonumber\\
 -&J_{z\pm}^{(3)}\Big( \Big( S^x_i S^z_j +S^z_i S^x_j \Big) c_\alpha 
 +\Big( S^y_i S^z_j+S^z_i S^y_j\Big)s_\alpha \Big), \nonumber
\end{align}
where $c_\alpha\equiv\cos\varphi_\alpha$ and $s_\alpha\equiv\sin\varphi_\alpha$ with the bond-dependent phases $\varphi_\alpha\!=\!\{0,2\pi/3,-2\pi/3\}$ for the three types of first and third-neighbor bonds ${\bm \delta}_\alpha$, $\bm{\delta}^{(3)}_\alpha$, shown in Fig.~\ref{fig_lattice_pd}(a). Note that this model is written relative to crystallographic $\{x,y,z\}$ axes shown in Fig.~\ref{fig_lattice_pd}(a), which are defined by the plane of the honeycomb lattice.  In Eq.~\eqref{eq_ham} we also include magnetic field parallel to $x$. The bond-independent terms in this paper are assumed to be of easy-plane $XY$ type: $\Delta_1=\Delta_3=0$. The bond-dependent term $J_{z\pm}^{(3)}$ is related to exchanges of the extended Kitaev-Heisenberg model
\begin{align}
\hat{\cal H}_\text{KH}=\sum_{\langle ij \rangle_n} \mathbf{S}_i^\intercal \hat{J}^{(n)}_\alpha \mathbf{S}_j,
~\hat{J}^{(n)}_\text{X}=
\begin{pmatrix}
\text{J}_n+K_n & \Gamma'_n & \Gamma'_n\\
\Gamma'_n & \text{J}_n & \Gamma_n\\
\Gamma'_n & \Gamma_n & \text{J}_n
\end{pmatrix},
\label{eq_ham_kh}
\end{align}
which is written relative to the cubic axes \{x,y,z\} defined by ion-ligand bonds, see Fig.~\ref{fig_lattice_pd}(a). There are three types of bonds $\alpha=\text{X,Y,Z}$, shown in Fig.~\ref{fig_lattice_pd}(a), for both $n=1$ nearest-neighbor and $n=3$ third-neighbor interactions. Interaction on X bond is given in Eq.~\eqref{eq_ham_kh}, the exchange matrix on Y and Z bonds is obtained through a cyclic permutation. Note that the models in crystallographic and cubic axes are equivalent and related to each other through a linear transformation \cite{rau2014trigonal,chaloupka2015,BCAO_minimal}.

\begin{figure}
\centering
\includegraphics[width=0.99\linewidth]{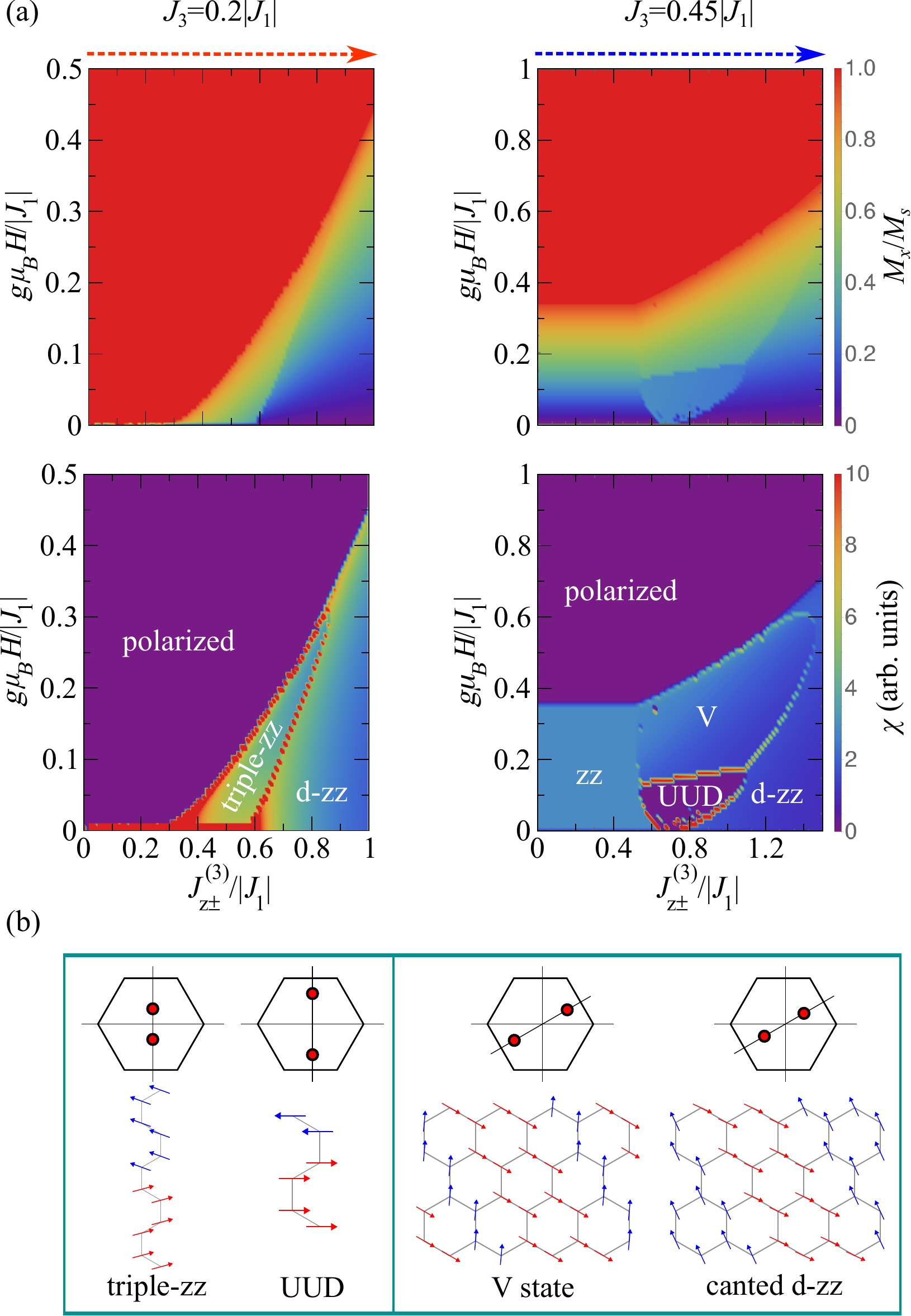}
\caption{(a) Intensity plots of magnetization and magnetic susceptibility for two representative values of $J_3$, shown as scans along $J_{z\pm}^{(3)}$, with phases indicated. (b) Sketches of the phases above with their ordering vectors shown in the BZ.}
\label{fig_scans}
\end{figure}

Phase diagram of the $J_1$-$J_3$-$J_{z\pm}^{(3)}$ model \eqref{eq_ham} for $J_1<0$ and $H=0$ is shown in Fig.~\ref{fig_lattice_pd}(c) and was calculated in Ref.~\cite{BCAO_minimal} in classical and quantum $S=1/2$ limits. In the classical limit, for $J_{z\pm}^{(3)}=0$ there is a transition from a ferromagnetic (FM) to zigzag (ZZ) state via intermediate spiral states, whose ordering vectors changes gradually from $\Gamma$ to $M$ \cite{Rastelli_1979}. The magnitude of the ordering vector of the phases in Fig.~\ref{fig_lattice_pd}(c) is indicated with color, and the corresponding colorbar is shown in Fig.~\ref{fig_lattice_pd}(b). When $J_{z\pm}^{(3)}$ is increased, double-zigzag state becomes stabilized in a wide region of phase diagram both in the classical limit and for $S=1/2$ \cite{BCAO_minimal}. Note that there are intermediate phases with various periodicity between FM and double-zigzag, such as $\mathbf{Q}=(1/6,0)$ ``triple-zigzag'', as well as $\mathbf{Q}=(1/3,0)$ UUD state between double-zigzag and ZZ. The structure of these states and their ordering vectors in the Brillouin zone are shown in Fig.~\ref{fig_scans}(b).

In order to explore field phase transitions and how these intermediate phases affect the field behavior, we perform two calculations for representative values of third-neighbor interaction, $J_3=0.2|J_1|$ and $J_3=0.45|J_1|$ (shown with dashed arrows in Fig.~\ref{fig_lattice_pd}(c)), where we plot phase diagram in the $J_{z\pm}^{(3)}-H$ axes. We perform numerical minimization of classical energy of Hamiltonian \eqref{eq_ham} on various clusters up to $12\times12$ sites using \texttt{Mathematica} software to identify the phases, with a subsequent minimization on smaller clusters to obtain phase boundaries. The result is shown in Fig.~\ref{fig_scans}(a) as intensity plot of magnetization $M_x$, and magnetic susceptibility $\chi=dM_x/dH$. 

For $J_3=0.2$ and small values of $J_{z\pm}^{(3)}$ there is a direct transition from a FM to polarized state. For $0.4<J_{z\pm}^{(3)}/|J_1|<0.6$ ground state is the canted triple-zigzag state, which further gradually cants towards saturation for $H>0$. At large $J_{z\pm}^{(3)}$ zero-field double-zigzag with three $C_3$-symmetric domains, selects only two domains of canted double-zigzag state, closest to the field direction. This canted double-zigzag state is illustrated in Fig.~\ref{fig_scans}(b). This situation is similar to the selection of zigzag domains in magnetic field in $\alpha$-RuCl$_3$ \cite{banerjee2018}, where three domains are degenerate in zero field but for $H>0$ only some are selected due to the presence of Kitaev terms \cite{sears2017}. Note that there is a region with  intermediate triple-zigzag state at high fields before double-zigzag reaches saturation.

For $J_3=0.45$ we observe even more intricate field behavior. First, for low values of $J_{z\pm}^{(3)}$, when the zero-field state is zigzag, as magnetic field is increased there is only canting of spins in the ZZ state towards the direction of the field without phase transition, which is illustrated by the absence of singularities in intensity plot of magnetic susceptibility $\chi$. As $J_{z\pm}^{(3)}$ increases, $\mathbf{Q}=(1/3,0)$ UUD phase becomes ground state, and it is presented as $1/3$-plateau as magnetic field is increased, until there is a transition to a canted ``V'' state at larger fields. Note that this state is analogous to the ``V'' state of triangular lattice antifferomagnet in magnetic field \cite{Kawamura_1984,Korshunov_1986,chubukov91}. For large enough $J_{z\pm}^{(3)}$, deep in the double-zigzag state, magnetic field only induces canting of spins towards the $x$ direction until saturation.

The most interesting situation occurs in the vicinity of transition between UUD and double-zigzag state around $J_{z\pm}^{(3)}=0.9|J_1|$, where UUD state extends its region of stability as magnetic field increases, which results in the transition from double-zigzag state to UUD state, similar to what is observed in experiments on BCAO \cite{Regnault_BaCoAsO_1977,Zhong2020,Armitage_THz_2021,Wang_THz_2021}. Notably, this transition  occurs only near the zero-field phase boundary between double-zigzag and UUD state. Therefore, the observation of this transition allows us to map out the region of parameters applicable to BCAO. However, classical analysis shows that at high field there is an intermediate ``V'' state before full polarization, which is not observed experimentally. We would like to argue that quantum fluctuations may be responsible for modifications to the field phase diagram.

{\it Quantum effects and parameters fit.-} Quantum fluctuations in the case of $S=1/2$ magnetic moments, such as the lowest doublet of BCAO, are known to strongly affect magnetic phase diagram compared to classical calculations -- $1/3$ plateau of triangular lattice antiferromagnet being a prime example \cite{chubukov91}. Since quantum calculations can be demanding, it was shown that their effect can be simulated by including biquadratic coupling to the classical model, which can be derived as first order of quantum fluctuations \cite{Henley_BQ,Starykh_plateau_11,expBaCoSbO2015,Zhitomirsky_2015} or used \textit{ad hoc} for various frustrated magnets \cite{chubukov91,Penc_plateau_04,Kaplan_plateau,Smirnov2017}. 

\begin{figure}
\centering
\includegraphics[width=0.99\linewidth]{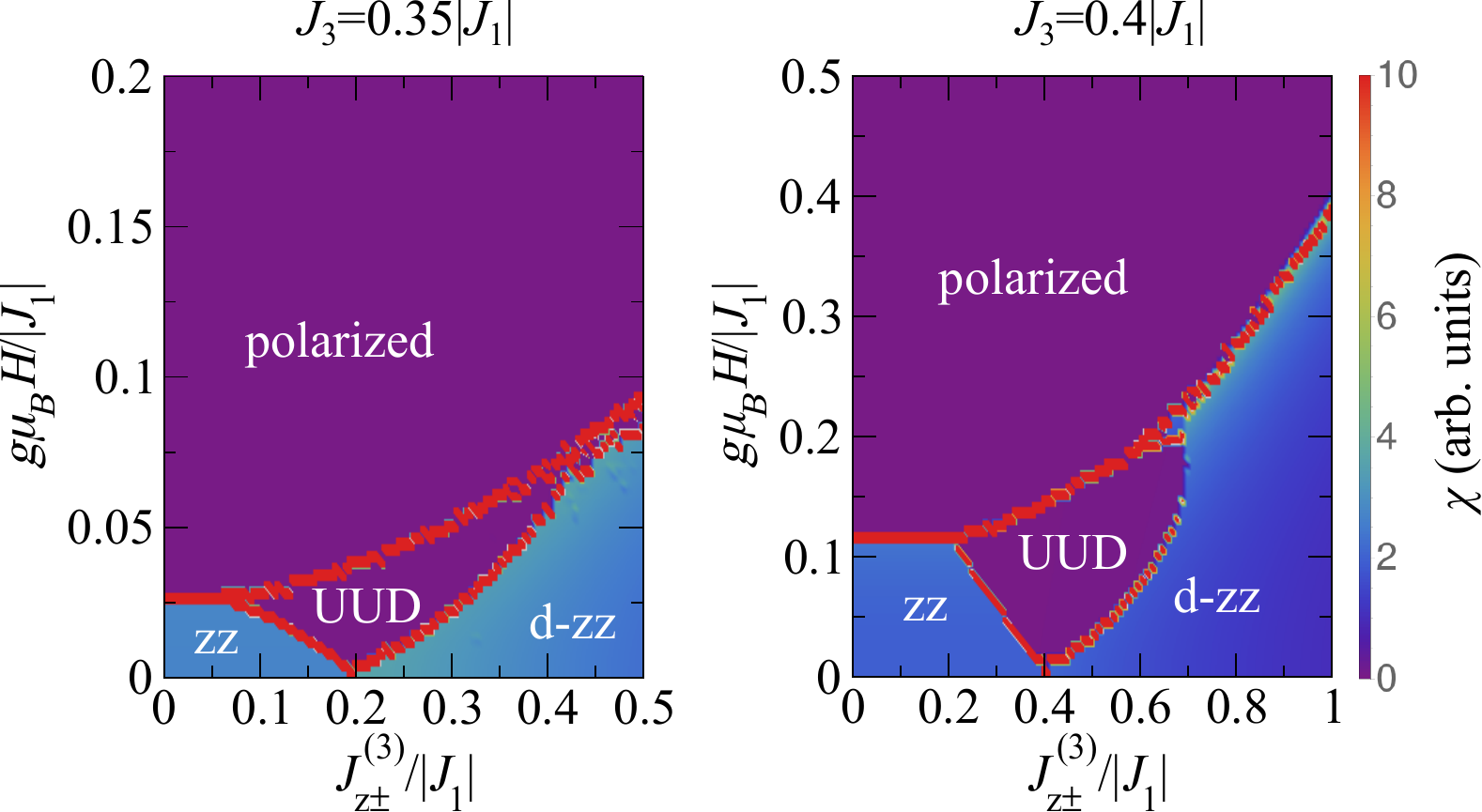}
\caption{Intensity plots of magnetic susceptibility for two representative values of third-neighbor interaction $J_3 \gtrsim 0.3|J_1|$ with phases indicated for the model \eqref{eq_ham} with biquadratic coupling \eqref{eq_bq} included, $B=0.04|J_1|$.}
\label{fig_fieldK}
\end{figure}

We also explore how quantum fluctuations, approximated by biquadratic coupling
\begin{align}
-B\sum_{\langle ij \rangle_1} \left(\mathbf{S}_i \mathbf{S}_j\right)^2,
\label{eq_bq}
\end{align}
affect the field phase diagram of model \eqref{eq_ham}. The results for $B=0.04|J_1|$ are shown in Fig.~\ref{fig_fieldK} for two representative values of $J_3$. This value is the lowest to suppress ``V'' state, and is of the same order as earlier estimates \cite{Starykh_plateau_11,expBaCoSbO2015,Zhitomirsky_2015}. One can see that collinear UUD state is stabilized in a wider region of parameters, while the ``V'' state is not present in the field phase diagram in Fig.~\ref{fig_fieldK} compared to Fig.~\ref{fig_scans}, leading to an experimentally observed double-zigzag$\rightarrow$UUD$\rightarrow$saturation sequence in a wide region of the phase diagram.

We have also checked other values of $B$ up to $0.12|J_1|$, and confirmed very weak changes of the results with the value of biquadratic coupling. This is due to the fact that zero-field and field-induced phases in the studied region of the phase diagram are nearly collinear. Therefore, biquadratic coupling is a constant, and does not affect the transitions after quantum effects are taken into account. Thus, in this paper we only present results for $B=0.04|J_1|$.

\begin{figure}
\centering
\includegraphics[width=0.99\linewidth]{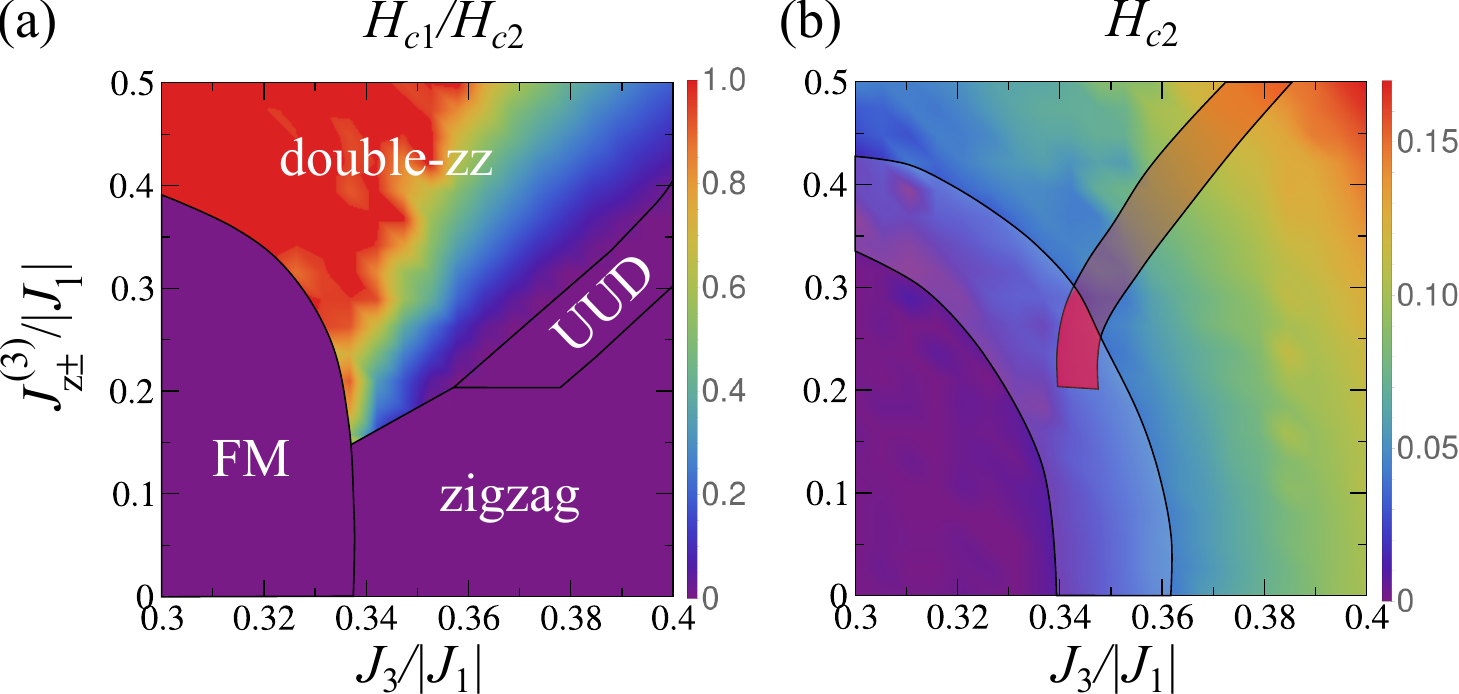}
\caption{(a) Intensity plot of $H_{c1}/H_{c2}$, two critical fields of the double-zigzag$\rightarrow$UUD$\rightarrow$saturated sequence of transitions. When $H_{c1}/H_{c2}=1$ there is only a direct transition from canted double-zigzag to polarized state. (b) Intensity plot of saturation field $H_{c2}$ in the units of $g \mu_B H/|J_1|$. Blue shaded region indicates parameters with $0.02<g \mu_B H_{c2}/|J_1|<0.04$. Red shaded region marks parameters with $0.4<H_{c1}/H_{c2}<0.6$. The intersection of these regions gives optimal parameter sets that fit experimentally observed values.}
\label{fig_hc_hs}
\end{figure}

Moreover, we can map out the region of the $J_3-J_{z\pm}^{(3)}$ phase diagram, where the values of critical magnetic fields fit experimental measurements, $H_{c1}$ and $H_{c2}$. Here, $H_{c1}$ is the critical of field of double-zigzag$\rightarrow$UUD transition, and $H_{c2}$ is the saturation field. According to magnetization measurements \cite{Regnault_BaCoAsO_1977,Zhong2020}, $H_{c1}=0.26T$, $H_{c2}=0.52T$. We break the parameter fit procedure into two steps. First, we plot $H_{c1}/H_{c2}$ as an intensity plot in the $J_3-J_{z\pm}^{(3)}$ phase diagram in Fig.~\ref{fig_hc_hs}(a), where one can see that such a sequence of field-induced transitions occurs only in the proximity of the UUD state. When $H_{c1}/H_{c2}$ reaches 1 in the phase diagram in Fig.~\ref{fig_hc_hs}(a), it implies the absence of intermediate UUD state and a direct transition from canted double-zigzag to polarized state. The range of parameters where $0.4<H_{c1}/H_{c2}<0.6$ is shown as the red shaded region in Fig.~\ref{fig_hc_hs}(b) fitting the experimentally observed ratio 0.5.

Next, we use $g=5.2$ \cite{Zhong2020} to map the region of phase diagram with saturation field $H_{c2}=0.52T$. Various experimental estimates of nearest-neighbor coupling range from $J_1=-3.3\text{ meV}$ \cite{Regnault_1986} to $J_1=-7.6\text{ meV}$ \cite{Broholm_BCAO} (see also Refs.\cite{Paramekanti_Co_2021,BCAO_minimal} for \textit{ab initio} calculations with similar results), which yields allowed region of values of $g \mu_B H_{c2}/|J_1|$ as 0.02--0.04. Parameters of Hamiltonian \eqref{eq_ham} which produce this range of critical fields are shown as the white shaded region in Fig.~\ref{fig_hc_hs}(b). Combined with the requirement for the ratio of critical fields $H_{c1}/H_{c2}$, we obtain the region of exchange parameters for BaCo$_2$(AsO$_4$)$_2$ that fit field-induced transitions observed in experiment, shown as the intersection of two shaded regions. We have also performed these calculations for the XXZ anisotropy $\Delta_1=\Delta_3=0.4$, which is motivated by spin-wave theory fit to the field-behavior of the $\mathbf{k}=0$ mode of BaCo$_2$(AsO$_4$)$_2$ in the polarized phase \cite{Broholm_BCAO}. The region of optimal parameters for $\Delta=0.4$ is very similar to the results in Fig.~\ref{fig_hc_hs}(b) due to the co-planar nature of the states in the phase diagram, which makes classical energies of the zero-field and field-induced states independent of choice of $\Delta$.

In order to illustrate the goodness of the fit, we plot magnetization for a representative set of parameters $J_{z\pm}^{(3)}=0.2|J_1|$, $J_3=0.34|J_1|$ in Fig.~\ref{fig_0}(b) together with experimental values from Ref.~\cite{Zhong2020} (we use $J_1=-7.6\text{ meV}$). There is a degree of discrepancy, which can be attributed to the broadening of transitions due to the finite temperature, while our calculations are performed at zero temperature. However, one can still see that our calculations provide very good fit for the experimental magnetization. Notably, the experimentally observed saturation field is rather small, and we see from Fig.~\ref{fig_hc_hs}(b) that the reason for that is the proximity of optimal parameter region to the region of multiple phases competition. Note that compared to previous estimate of exchanges in BCAO \cite{BCAO_minimal}, this parameter set requires larger $\Gamma_1$ and $\Gamma'_1$ - due to stronger easy-plane anisotropy, as well as larger $K_3$, $\Gamma_3$, $\Gamma_3'$ to describe the presence of non-negligible $J_{z\pm}^{(3)}$. 

\begin{figure}
\centering
\includegraphics[width=0.75\linewidth]{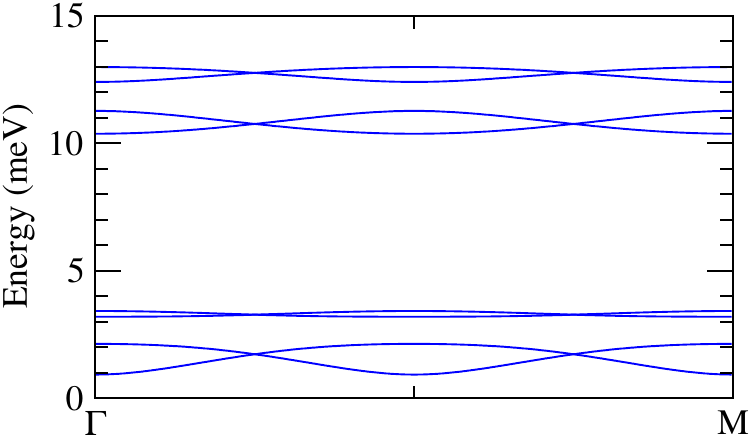}
\caption{Spin-wave magnetic spectrum in the double-zigzag state for $J_1=-7.6\text{ meV}$, $J_{z\pm}^{(3)}=0.4|J_1|$, $J_3=0.34|J_1|$.}
\label{fig_spectrum}
\end{figure}

Moreover, in order to check that a rather large bond-dependent term $J_{z\pm}^{(3)}$,  which is necessary for explanation of the features studied in this paper, does not introduce anomalies to the magnetic spectrum, we have performed linear spin-wave calculations for $J_{z\pm}^{(3)}=0.4|J_1|$, $J_3=0.34|J_1|$, $J_1=-7.6\text{ meV}$. The result is shown in Fig.~\ref{fig_spectrum} with energy in units of meV for a direct comparison with inelastic neutron scattering data \cite{dejongh1990,Broholm_BCAO}. While we cannot expect a complete agreement with experimental data, we see that our calculation does reproduce multiple features of the spectrum. For instance, there are low-energy modes with a small gap around 1 meV (1.5 meV experimentally), as well as high-energy modes around 11 and 13 meV, in agreement with a 12 meV mode in neutron scattering data \cite{dejongh1990}.  Moreover, there is a rather flat mode around 3.5 meV in Fig.~\ref{fig_spectrum}, which was also observed in neutron scattering data \cite{Broholm_BCAO}. Note that even in the presence of large $J_{z\pm}^{(3)}=3\text{ meV}$, the gap is rather small, which means that the model proposed in this paper is physically viable. Nonetheless, we should note that the spin-wave energies in Fig.~\ref{fig_spectrum} do not give precise agreement with neutron scattering data and a careful fitting of magnetic spectrum assumably requires all eight parameters of the extended Kitaev-Heisenberg model, placing such fit beyond the scope of this work.

{\it Discussion.-} Establishing parameters of a model with multiple allowed exchange parameters, like extended Kitaev-Heisenberg model, can become cumbersome in a lot of cases. Experiments in magnetic field, which can include several transitions, may strongly assist in setting boundaries on exchanges. We performed analysis of field-induced phases of the anisotropic exchange model suggested previously for BaCo$_2$(AsO$_4$)$_2$ using numerical minimization of classical spins, and we observed a plethora of phase transitions in various parts of the zero-field phase diagram. We showed that experimentally established double-zigzag$\rightarrow$UUD$\rightarrow$saturated phase sequence is stabilized by quantum fluctuations and only in the proximity of the double-zigzag phase boundary. By scanning through the phase space, we fit critical fields to the experimentally observed values, and narrow down the parameters which provide the best agreement. As it turns out, the presence of third-neighbor bond-dependent exchange $J_{z\pm}^{(3)}$ is necessary, and in our analysis we are able to put limits on its values. We should note that previous attempts at calculating exchanges for BaCo$_2$(AsO$_4$)$_2$ are not in a precise agreement with each other. However, they all point to an easy-plane $XXZ$ $J_1$-$J_3$ model, with the ratio of nearest- and third-neighbor exchanges around 0.3, and the presence of small but non-negligible bond-dependent corrections \cite{Regnault_1986,Paramekanti_Co_2021,BCAO_minimal,Broholm_BCAO}. Our parameter set supports this notion and further promotes the idea that third-neighbor bond-dependent term is crucial for the physics of BaCo$_2$(AsO$_4$)$_2$. Moreover, recent measurements on a related compound BaCo$_2$(PO$_4$)$_2$ indicate a very similar field-induced behavior \cite{BaCoPO}, making our results applicable for a broad class of materials.

{\it Acknowledgments.} I would like to thank Sergey Streltsov and Sasha Chernyshev for useful discussions. I acknowledge support of the Russian Science Foundation via project 23-12-00159.
\bibliography{jpp_bib}
\end{document}